\documentstyle[epsf]{article}

\catcode`\@=11
\long\def\@makefntext#1{
\protect\noindent \hbox to 3.2pt {\hskip-.9pt
$^{{\eightrm\@thefnmark}}$\hfil}#1\hfill}       

\def\@makefnmark{\hbox to 0pt{$^{\@thefnmark}$\hss}}    

\def\ps@myheadings{\let\@mkboth\@gobbletwo
\def\@oddhead{\hbox{}
\rightmark\hfil\eightrm\thepage}
\def\@oddfoot{}\def\@evenhead{\eightrm\thepage\hfil
\leftmark\hbox{}}\def\@evenfoot{}
\def\sectionmark##1{}\def\subsectionmark##1{}}

\oddsidemargin=\evensidemargin
\newcommand{\us}{|\!\!\uparrow\rangle}
\newcommand{\ds}{|\!\!\downarrow\rangle}

\newcounter{sectionc}\newcounter{subsectionc}\newcounter{subsubsectionc}
\renewcommand{\section}[1] {\vspace{12pt}\addtocounter{sectionc}{1}
\setcounter{subsectionc}{0}\setcounter{subsubsectionc}{0}\noindent
    {\tenbf\thesectionc. #1}\par\vspace{5pt}}
\renewcommand{\subsection}[1] {\vspace{12pt}\addtocounter{subsectionc}{1}
\setcounter{subsubsectionc}{0}\noindent
{\bf\thesectionc.\thesubsectionc. {\kern1pt \bfit #1}}\par\vspace{5pt}}
\renewcommand{\subsubsection}[1] {\vspace{12pt}\addtocounter{subsubsectionc}{1}
    \noindent{\tenrm\thesectionc.\thesubsectionc.\thesubsubsectionc.
    {\kern1pt \tenit #1}}\par\vspace{5pt}}
\newcommand{\nonumsection}[1] {\vspace{12pt}\noindent{\tenbf #1}
    \par\vspace{5pt}}

\newcounter{appendixc}
\newcounter{subappendixc}[appendixc]
\newcounter{subsubappendixc}[subappendixc]
\renewcommand{\thesubappendixc}{\Alph{appendixc}.\arabic{subappendixc}}
\renewcommand{\thesubsubappendixc}
    {\Alph{appendixc}.\arabic{subappendixc}.\arabic{subsubappendixc}}

\renewcommand{\appendix}[1] {\vspace{12pt}
        \refstepcounter{appendixc}
        \setcounter{figure}{0}
        \setcounter{table}{0}
        \setcounter{lemma}{0}
        \setcounter{theorem}{0}
        \setcounter{corollary}{0}
        \setcounter{definition}{0}
        \setcounter{equation}{0}
        \renewcommand{\thefigure}{\Alph{appendixc}.\arabic{figure}}
        \renewcommand{\thetable}{\Alph{appendixc}.\arabic{table}}
        \renewcommand{\theappendixc}{\Alph{appendixc}}
        \renewcommand{\thelemma}{\Alph{appendixc}.\arabic{lemma}}
        \renewcommand{\thetheorem}{\Alph{appendixc}.\arabic{theorem}}
        \renewcommand{\thedefinition}{\Alph{appendixc}.\arabic{definition}}
        \renewcommand{\thecorollary}{\Alph{appendixc}.\arabic{corollary}}
        \renewcommand{\theequation}{\Alph{appendixc}.\arabic{equation}}
        \noindent{\tenbf Appendix \theappendixc #1}\par\vspace{5pt}}
\newcommand{\subappendix}[1] {\vspace{12pt}
        \refstepcounter{subappendixc}
        \noindent{\bf Appendix \thesubappendixc. {\kern1pt \bfit #1}}
    \par\vspace{5pt}}
\newcommand{\subsubappendix}[1] {\vspace{12pt}
        \refstepcounter{subsubappendixc}
        \noindent{\rm Appendix \thesubsubappendixc. {\kern1pt \tenit #1}}
    \par\vspace{5pt}}

\newcommand{\textlineskip}{\baselineskip=13pt}
\newcommand{\smalllineskip}{\baselineskip=10pt}

\newcommand{\copyrightheading}[1]
    {\vspace*{-2.5cm}\smalllineskip{\flushleft
    {\footnotesize Quantum Information and Computation, Vol.~1, No.~0 (2001) 000--000 #1}\\
    {\footnotesize \copyright\kern2pt Rinton Press}\\
     }}

\newcommand{\publisher}[2]{{\begin{center}\footnotesize\smalllineskip
    Received #1\\
    Revised #2
    \end{center}
    }}

\def\abstracts#1#2#3{{
    \centering{\begin{minipage}{4.5in}\footnotesize\baselineskip=10pt
    \parindent=0pt #1\par
    \parindent=15pt #2\par
    \parindent=15pt #3
    \end{minipage}}\par}}

\def\keywords#1{{
    \centering{\begin{minipage}{4.5in}\footnotesize\baselineskip=10pt
    {\footnotesize\it Keywords}\/: #1
     \end{minipage}}\par}}
\def\communicate#1{{
    \centering{\begin{minipage}{4.5in}\footnotesize\baselineskip=10pt
    {\footnotesize\it Communicated by}\/: #1
     \end{minipage}}\par}}

\renewenvironment{thebibliography}[1]
        {\frenchspacing
     \ninerm\baselineskip=11pt
         \begin{list}{\arabic{enumi}.}
        {\usecounter{enumi}\setlength{\parsep}{0pt}
     \setlength{\leftmargin 12.7pt}{\rightmargin 0pt}
         \setlength{\itemsep}{0pt} \settowidth
    {\labelwidth}{#1.}\sloppy}}{\end{list}}

\newcounter{itemlistc}
\newcounter{romanlistc}
\newcounter{alphlistc}
\newcounter{arabiclistc}

\newcommand{\fcaption}[1]{
        \refstepcounter{figure}
        \setbox\@tempboxa = \hbox{\footnotesize Fig.~\thefigure. #1}
        \ifdim \wd\@tempboxa > 5in
           {\begin{center}
        \parbox{5in}{\footnotesize\smalllineskip Fig.~\thefigure. #1}
            \end{center}}
        \else
             {\begin{center}
             {\footnotesize Fig.~\thefigure. #1}
              \end{center}}
        \fi}

\newcommand{\tcaption}[1]{
        \refstepcounter{table}
        \setbox\@tempboxa = \hbox{\footnotesize Table~\thetable. #1}
        \ifdim \wd\@tempboxa > 5in
           {\begin{center}
        \parbox{5in}{\footnotesize\smalllineskip Table~\thetable. #1}
            \end{center}}
        \else
             {\begin{center}
             {\footnotesize Table~\thetable. #1}
              \end{center}}
        \fi}

\def\pmb#1{\setbox0=\hbox{#1}
    \kern-.025em\copy0\kern-\wd0
    \kern.05em\copy0\kern-\wd0
    \kern-.025em\raise.0433em\box0}

\def\fnt#1#2{\footnotetext{\kern-.3em
    {$^{\mbox{\scriptsize #1}}$}{#2}}}

\def\fpage#1{\begingroup
\voffset=.3in
\thispagestyle{empty}\begin{table}[b]\centerline{\footnotesize #1}
    \end{table}\endgroup}

\def\runninghead#1#2{\pagestyle{myheadings}
\markboth{{\protect\footnotesize\it{\quad #1}}\hfill}
{\hfill{\protect\footnotesize\it{#2\quad}}}}
\font\tenrm=cmr10
\font\tenit=cmti10
\font\tenbf=cmbx10
\font\bfit=cmbxti10 at 10pt
\font\ninerm=cmr9

\font\eightrm=cmr8

\def\FigName{figure}%
\newbox\captionbox
\long\def\@makecaption#1#2{%
  \ifx\FigName\@captype
    \vskip\abovecaptionskip
    \setbox\tempbox\hbox{{\figurecaptionfont #1\hskip1em #2}}
    \ifdim\wd\tempbox< 28pc
    \centerline{\box\tempbox}
    \else
    {\figurecaptionfont #1\hskip1em #2\par}
\fi\else
    \setbox\tempbox\hbox{{\tablecaptionfont #1\hskip1em #2}}
    \ifdim\wd\tempbox< 28pc
    \centerline{\box\tempbox}
    \else
    {\tablecaptionfont #1\hskip1em #2\par}%
    \fi
 \vskip\belowcaptionskip
 \fi}
\InputIfFileExists{psfig.sty}
{\typeout{^^Jpsfig.sty inputed...ok}}{\typeout{^^JWarning: psfig.sty could be be found.^^J}}
\InputIfFileExists{epsfsafe.tex}
{\typeout{^^Jepsfsafe.tex inputed...ok}}
            {\typeout{^^JWarning: epsfsafe.tex could not be found.^^J}}
\InputIfFileExists{epsfig.sty}
{\typeout{^^Jepsfig.sty inputed...ok}}{\typeout{^^JWarning: epsfig.sty could not be found.^^J}}
\InputIfFileExists{epsf.sty}
{\typeout{^^Jepsf.sty inputed...ok}}{\typeout{^^JWarning: epsf.sty could not be found.^^J}}%
\def\fps@figure{tbp}
\def\ftype@figure{1}
\def\ext@figure{lof}
\def\fnum@figure{Fig.\ \thefigure}
\def\qed{\hbox{${\vcenter{\vbox{              
   \hrule height 0.4pt\hbox{\vrule width 0.4pt height 6pt
   \kern5pt\vrule width 0.4pt}\hrule height 0.4pt}}}$}}

\begin{document}
\setlength{\textheight}{8.0truein}    

\runninghead{Transport of Quantum States   $\ldots$}
            {M. A. Rowe, \emph{et al.}}

\normalsize\textlineskip
\thispagestyle{empty}
\setcounter{page}{1}

\copyrightheading{} 

\vspace*{0.88truein}

\fpage{1}
\centerline{\bf
TRANSPORT OF QUANTUM STATES AND SEPARATION OF IONS}
\vspace*{0.035truein} \centerline{\bf IN A DUAL RF ION
TRAP\footnote{Contribution of the National Institute of Standards
and Technology; not subject to U.S. copyright.}}
\vspace*{0.37truein} \centerline{\footnotesize}
\centerline{M.~A.~ROWE\footnote{present address: Optoelectronics
Division, NIST Boulder.}, A.~BEN-KISH\footnote{present address:
Technion, Haifa, Israel}, B.~DEMARCO, D.~LEIBFRIED, V.~MEYER,}
\centerline{J. BEALL\footnote{Electromagnetic Technology Division,
NIST Boulder}, J.~BRITTON, J.~HUGHES, W.~M.~ITANO,
B.~JELENKOVI\'{C},} \vspace*{0.015truein} \centerline{ C.~LANGER,
T.~ROSENBAND, and D.~J.~WINELAND} \vspace*{0.015truein}
\centerline{\footnotesize\it Time and Frequency Division, National
Institute of Standards and Technology,} \baselineskip=10pt
\centerline{\footnotesize\it 325 S. Broadway, Boulder, Colorado
80305-3328} \vspace*{0.225truein} \publisher{(received
date)}{(revised date)}

\vspace*{0.21truein}
\abstracts{
We have investigated ion dynamics associated with a dual linear
ion trap where ions can be stored in and moved between two
distinct locations.  Such a trap is a building block for a system
to engineer arbitrary quantum states of ion ensembles.
Specifically, this trap is the unit cell in a strategy for
scalable quantum computing using a series of interconnected ion
traps.  We have transferred an ion between trap locations 1.2 mm
apart in 50 $\mu$s with near unit efficiency ($> 10^{6}$
consecutive transfers) and negligible motional heating, while
maintaining internal-state coherence.  In addition, we have
separated two ions held in a common trap into two distinct
traps.}{}{}

\vspace*{10pt} \keywords{atom trapping and cooling, laser cooling,
quantum computation, quantum state engineering, trapped ions}
\vspace*{3pt} \communicate{to be filled by the Editorial}

\vspace*{1pt}\textlineskip  
\section{Introduction}          
\vspace*{-0.5pt}
\noindent
Developing a practical method for synthesizing and manipulating
entangled quantum states of many particles is challenging.  One
approach is to physically isolate small subsystems that can be
operated on independently.  By entangling these subsystems
sequentially or in parallel a large entangled state can be
constructed \cite{lloyd93}. Some of the elements required for this
approach, in the context of trapped ions, are demonstrated here.

For quantum information processing, it may be desirable to
transport quantum states between nodes of a larger system for
various technical or intrinsic reasons (e.g., in communication
systems).  In a quantum computer, the carriers might be photons
\cite{cirac97} \nocite{pellizzari97} - \cite{deVoe98}, electron
spins \cite{kikkawa99}\nocite{barnes00} - \cite{recher01}, atomic
ions \cite{wineland98,cirac00}, or neutral atoms that are
transported with guides \cite{renn95}
\nocite{denschlag99,muller99,dekker00,fortagh00,key00,cassettari00}
- \cite{muller01} or movable potentials \cite{reichel99}
\nocite{schrader01} - \cite{gustavson02}. For ions, this basic
idea has also been applied to atomic clocks \cite{prestage94} and
cavity-QED studies \cite{guthohrlein01}.

There are stringent requirements for a quantum computing system
where perhaps hundreds or thousands of two-level quantum systems
(qubits) interact.  Cirac and Zoller \cite{cirac95} proposed a
physical scheme to implement quantum computing using a string of
ions that are held in a single linear RF trap.  Two internal
states of each ion form a qubit.  An applied focused laser field
selectively couples the qubits to a shared quantized vibrational
mode of the trap. Gate operations between qubits are implemented
by way of this common vibrational mode that acts as a data bus.

The Cirac-Zoller scheme provides a good starting point for quantum
operations, but as the number of ions increases several
difficulties are encountered.  The addition of each ion to the
string adds three vibrational modes.  It soon becomes nearly
impossible to spectrally isolate the desired vibrational mode
unless the speed of operations is slowed to undesirable levels
\cite{wineland98,steane00}.  In addition, as ions are added to the
trap the axial trap strength must be decreased in order to
maintain a linear string of ions \cite{enzer00}. A weaker trap
makes sideband laser cooling less efficient and aggravates the
problem of mode isolation due to multimode excitations
\cite{wineland98}. Furthermore, since error correction will most
likely be incorporated into any large processor, it will be
desirable to measure and reset ancilla qubits without disturbing
the coherence of logical qubits.  Since ion qubits are typically
read out using laser scattering, the scattered light from ancilla
qubits held in a common trap may disturb the coherence of the
logical qubits.

For these reasons, we have considered an architecture that employs
an array of interconnected ion traps
\cite{wineland98,kielpinski02}. Ions are moved between nodes in
the array by applying time-dependent potentials to ``control"
electrode segments.  To perform logic operations between selected
ions, these ions are transferred into an ``accumulator" trap for
the gate operation. Before the gate operation is performed, it may
be necessary to sympathetically re-cool the qubit ions with
another ion species \cite{wineland98}.  Subsequently, these ions
are moved to memory locations or other accumulators. This strategy
always maintains a relatively small number of motional modes that
must be considered and minimizes the problems of ion/laser-beam
addressing using focused laser beams. Such arrays also enable
highly parallel processing and ancilla qubit readout in a separate
trapping region so that the logical ions are shielded from the
scattered laser light. In the work reported here, we have
implemented the first steps towards realizing this architecture.
By continually changing the electric potentials on a series of
electrode segments, we were able to smoothly translate our
trapping potential and move ions adiabatically between two
locations in a simple ``dual-trap" array. In addition, by
increasing the potential on an electrode near two ions, a
potential wedge was inserted between the ions, separating them
into the two traps.

\section{The Trap}
\noindent The trap was constructed from a stack of metallized 200
$\mu$m thick alumina wafers. Laser-machined slots and gold traces
created the desired electrode geometry as in the traps of Ref.
\cite{turchette00}. Gold traces of 0.5 $\mu$m thickness were made
with evaporated gold that was transmitted through a shadow mask
and deposited on the alumina. Subsequently, an additional 3 $\mu$m
of gold was electroplated onto the electrodes, resulting in
electrode surfaces smooth at the 1 $\mu$m level. These
lithographic techniques allow for small traps and could be
expandable to larger arrays.  The idealized four-rod linear trap
geometry (Fig. 1a) is approximated using a wafer stack (Figs. 1b
and 1c).
\begin{figure} 
\vspace*{13pt}
\centerline{\psfig{file=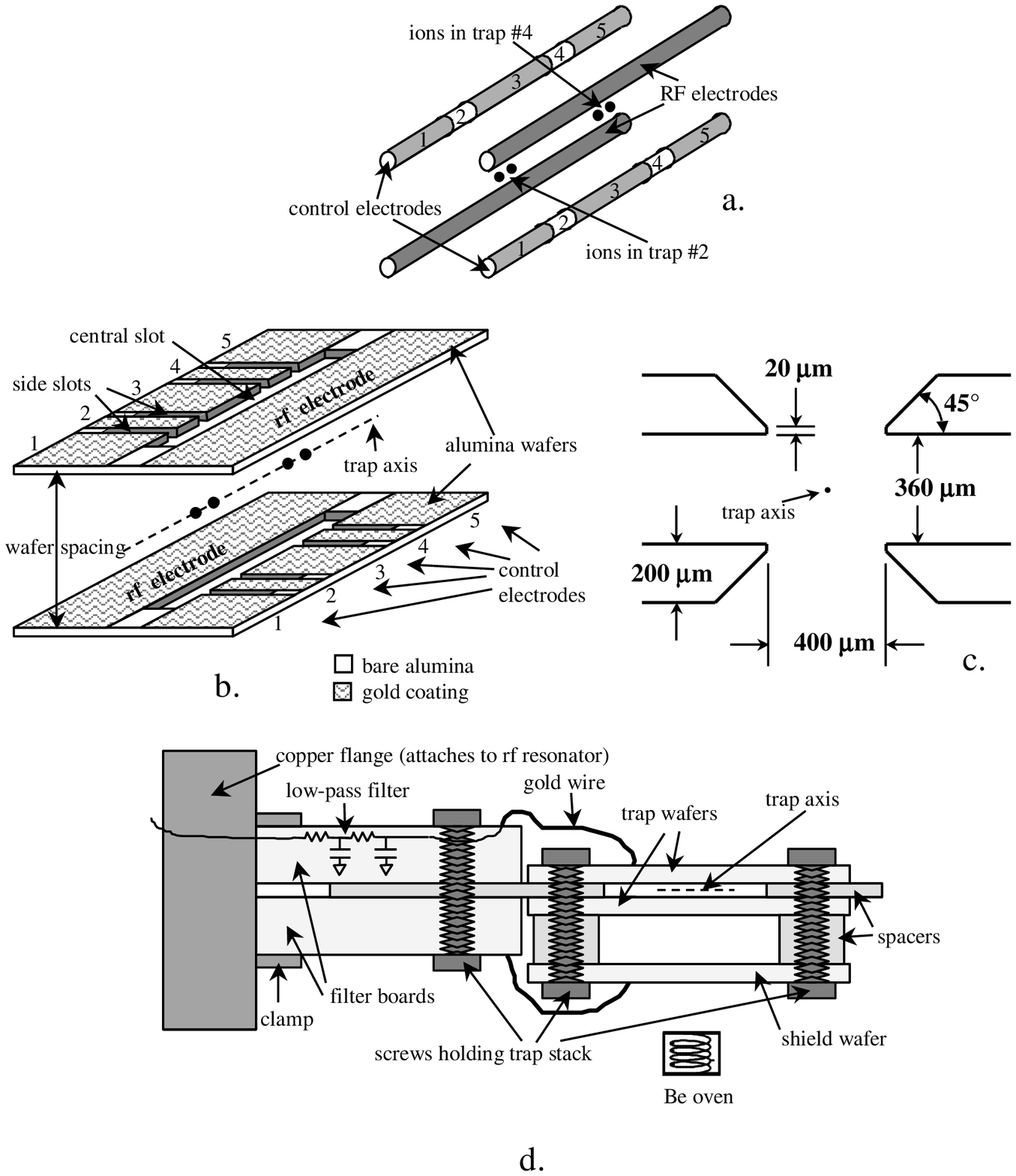, width=13cm}} 
\vspace*{13pt} \fcaption{\label{fig_trap} The dual linear ion trap
(drawings not to scale).  a. The idealized four-rod geometry. b.
The wafer-stack implementation.  The two trap wafers are spaced
with two 360 $\mu$m thick alumina pieces (not shown) that are
placed between them along the short edges.  The pairs of control
electrodes are numbered 1 through 5 for reference. The two trap
locations, $\#$2 and $\#$4, shown in the figure are labeled by the
electrode on which they are centered. The axial length of
electrode 1 (2,3,4,5) is 1100 $\mu$m (400 $\mu$m, 800 $\mu$m, 400
$\mu$m, 1100 $\mu$m).  For 8.0 V applied to electrodes 1, 3, and 5
and 0.0 V applied to electrodes 2 and 4 the axial trap frequency
in each trap was 2.9 MHz for a single $^{9}$Be$^{+}$ ion.  The
peak amplitude of the applied RF voltage was about 500 V. The RF
drive frequency was 230 MHz.  c. Cross-section of the trap
electrodes (looking along the trap axis). d. Side view assembly
diagram of the trap structure.}
\end{figure}
The central slot width (400 $\mu$m) and the wafer spacing (360
$\mu$m) set the length scale of the RF trap.  The side slots (10
$\mu$m wide) electrically isolate the different control
electrodes.

The control electrodes (held at RF ground) are arranged into five
segments for axial confinement (Fig. 1b). Radial confinement
(perpendicular to the axial direction) is provided primarily by
ponderomotive forces generated by potentials applied to the RF
electrodes \cite{wineland98}. Ions can be trapped at various
locations along the axis.  For instance, confinement (of positive
ions) at two distinct places occurs when the outer two and middle
segments are biased above the remaining two segments as shown
schematically in Fig. 1a.

A new feature of this trap compared to our previous ones is the
line-of-sight shielding of the electrodes from the neutral
beryllium oven source (located about 2 cm from the trap). Previous
evidence \cite{turchette00} indicated that the heating rate of an
ion's motional state increased over time as beryllium was
deposited onto the electrodes. In all of our experiments, the
source of beryllium has been a tungsten/beryllium filament that
was heated with direct current thereby emitting beryllium atoms
towards the trap electrodes. Some of the beryllium atoms that pass
through the trap are ionized by electrons that are also directed
through the trap; the resulting $^9$Be$^+$ ions are then confined.
During trap loading the beryllium is emitted in all directions and
in previous designs was able to coat the electrodes. In the
current trap, the electrodes were shielded from this beryllium
flux by a shadow mask. This consisted of an alumina wafer with a
single 200 $\mu$m wide slot cut along its center line that was
positioned 700 $\mu$m above the trap wafer nearest to the oven.
This shield wafer prevented deposition of beryllium on any part of
the trap electrodes. Gold was deposited onto a rectangular area
centered around the slot in the shield and facing the trap
electrodes. Voltage applied to the shield's gold trace in
combination with a differential voltage applied between diagonal
pairs of the control electrodes (for example, between the $\#$2
electrodes in trap $\#$2) allowed us to compensate for micromotion
in at least one of the traps \cite{berkeland98}.

Figure 1d shows a schematic of the entire trap structure. The two
trap wafers and the shield wafer were spaced with 360 $\mu$m thick
alumina wafers along the short edges.  The central and side slots
were unobstructed by these spacers.  The entire stack of wafers
was held together with two 000-120 screws (one on each end of the
wafer stack) passing through the trap axis. Two ``filter boards"
held the wafer stack via a mutual spacer board and an additional
screw. A cascade of two RC low-pass filters (R= 1 k$\Omega$, C=
820 pF) was connected to each control electrode by a gold wire.
The trap stack was clamped inside a resonator that applied up to
about 500 V at 230 MHz to the RF electrodes \cite{jefferts95},
which were electrically common. This trap assembly was encased in
a quartz envelope attached to a stainless steel vacuum chamber.
After cleaning, the system was baked at 300$^{\circ}$ C for about
two days, resulting in a final ambient pressure that was less than
$4\times 10^{-9}$ Pa ($3\times 10^{-11}$ Torr).

\section{The Apparatus}
\noindent We confine $^9$Be$^+$ ion qubits in this trap.  We
spectrally resolve two of the ions' ground-state hyperfine levels,
$F=2,m_{F}=-2$ and $F=1,m_{F}=-1$ whose wave functions are labeled
$\ds$ and $\us$ respectively.  The frequency splitting between
these two levels, $\omega_{0}/2\pi$, is approximately 1.25 GHz. We
must also account for the quantized motional levels due to the
ions' confinement by the trap potential. For a given motional
mode, these realize a ladder of states labeled $\mid \!\! n
\rangle$ where $n=0,1,2, \rightarrow \infty $.  We cool to the
ground state ($n=0$) with Doppler cooling followed by Raman
sideband cooling \cite{monroe95,king98}. ``Repump'' lasers
optically pump an ion to $\ds$, thereby initializing the internal
state \cite{monroe95}. We drive stimulated-Raman transitions with
laser beams to coherently move population between the atomic and
motional levels \cite{meekhof96,sackett01}.  The two Raman beams
have a wavelength $\lambda \simeq$  313 nm and a difference
frequency near $\omega_{0}/2\pi$. They are aligned either parallel
(0$^{\circ}$ geometry) or perpendicular (90$^{\circ}$ geometry) to
each other, with their difference wave vectors, $\vec{\Delta
k}\simeq 0$ or $|\vec{\Delta k}| \simeq \frac{2 \sqrt{2}
\pi}{\lambda}$ along the trap axis respectively.  The final state
is determined by probing the ion with circularly polarized light
from a ``detection'' laser. During the 200 $\mu$s detection
period, ions in the $\ds$ state scatter many photons, while ions
in the $\us$ state scatter few photons.  The total number of
fluorescence photons collected during the detection period
indicate the final spin state \cite{sackett01,rowe01}.

\section{Ambient Ion Heating}
\noindent The ambient heating of the motional modes is an
important source of decoherence for the Cirac/Zoller gates, which
proceed through specific vibrational levels \cite{cirac95}.  An
expected source of heating is that caused by thermal electronic
noise \cite{wineland98}.  However, the heating that we have
observed has always been significantly above what is expected from
this source and may be due to fluctuating patch potentials on the
electrode surfaces \cite{turchette00}. Compared to our previous
results \cite{turchette00}, the results for the dual trap given
below show that the ambient heating has been significantly
reduced\footnote{\ \ Heating results from several trapped ion
experiments are summarized in Refs. \cite{turchette00} and
\cite{Rohde01}. Here we limit our comparison to $^9$Be$^+$ heating
in traps with very similar construction.}.  Our heating figure of
merit, the time interval to absorb one motional quantum from the
ground state ($\approx$ 10 ms) divided by the entangling gate
period ($\approx$ 10 $\mu$s), approaches 1000. The information
below compares the results from three different traps and is
summarized in Table 1.
\begin{table}
\label{heat_summary} \tcaption{Summary of ambient heating results
for three different traps.  The heating rates are expressed as 1
quantum absorbed per unit of time and $\nu_z$ is the trap axial
frequency.} \centerline{\footnotesize }
\centerline{\footnotesize\smalllineskip
\begin{tabular}{c c c c c c c }\\
\hline
Trap & heating rate                                               &heating rate           &edge       &electro- &electrode    &electrode  \\
            &($\nu_z =$ 2.9 MHz)                                    &($\nu_z =$ 3.9 MHz)             &  plating       & plated   &shield      &distance ($d$)  \\
\hline
dual     &1/10 ms (initial)                   &  &yes            &yes                 &yes            &270 $\mu$m  \\
            &1/4 ms (later)                       &1/8 ms                            &                 &                      &                 &                       \\
trap A  &$\approx$ 1/100 $\mu$s      &$\approx$ 1/100 $\mu$s  &no     &no       &no       &220 $\mu$m  \\
trap C  &---                                           &1/4 ms (initial)   &yes               &no                 &no          &220 $\mu$m  \\
            &                                                & 1/1 ms (later)   &                    &                      &                 &                       \\
\hline\\
\end{tabular}}
\end{table}

We determined the ambient motional heating rate for the axial mode
(frequency $\nu_z$) of a single ion in the trap.  The ion's axial
motion is first cooled to the ground state.  Following this, the
thermal motional state, quantified by the mean populated motional
level $\langle n_z \rangle$, was measured for various delay times
by use of a standard sideband comparison technique
\cite{turchette00,monroe95}.  The heating rate, $\partial{\langle
n_z \rangle}/\partial t$, was initially measured to be 1
quantum/10 ms for $\nu_z =$ 2.9 MHz. This rate is about 100 times
smaller than in a previous trap\footnote{\ \ Trap 6 in Ref.
\cite{turchette00}.} \ (referred to here as trap A), which was
made with similar construction (evaporated gold on alumina wafers
but no electroplating). The characteristic distance $d$, which we
take as the distance from the ion to the nearest trap electrode,
is about $25\ \%$ larger for the dual trap than trap A\footnote{\
\ Some of the sizes, $d$, listed in Table I of Ref.
\cite{turchette00} are incorrect. The size for traps 4 and 5
should be 160 $\mu$m. The size for trap 6 (referred to here as
trap A) should be 220 $\mu$m.}. \ \ The approximate $d^{-4}$
scaling of the heating rate found in Ref. \cite{turchette00}
predicts that the heating rate in the dual trap should be smaller
by only about a factor of 2 than in trap A. The heating rate in
the dual trap increased to approximately 1 quantum/4 ms after a
period of about 2 weeks, but the rate stabilized to this value for
the next 6 months or about 1000 trap loads.

The line-of-sight shielding of the electrodes from beryllium
deposition appears to explain some of the improvement.  However,
 measurements of heating rates for another recently-constructed trap
(referred to here as trap C) indicate that beryllium contamination
may not be the only source of heating.  Trap C was a duplicate of
trap A (about the same dimensions and no electrode shielding)
except that additional care was taken with the electrode surfaces.
For both trap C and the dual trap we made sure that the evaporated
gold layer was even and continuous over the edges of all of the
electrodes nearest to the ions.  The  initial heating rate in trap
C (when the beryllium deposition was minimal) was 1 quantum per 4
ms for $\nu_z =$ 3.9 MHz. This rate is also much smaller (about a
factor of 35) than the rate in trap A with $\nu_z =$ 3.9 MHz.
However, after about 100 loads (the extent of our data) the
heating rate in trap C increased to about 1 quantum per 1 ms. The
heating rate in the dual trap with $\nu_z =$ 3.9 MHz was 1 quanta
per 8 ms (after 2 weeks of operation). The lower heating rate of
the dual trap relative to the initial heating rate in trap C could
be explained by the electroplating of the dual trap's electrodes
(the electrodes of traps A and C were not electroplated), some
initial beryllium contamination on trap C (before the first
heating measurements were carried out), the size scaling ($\approx
d^{-4}$), or some combination of these factors. Nevertheless, from
these results, electrode surface purity and smoothness appear to
be closely correlated with ion heating.

We can estimate the heating from thermal electronic noise in the
following way \cite{wineland98}: Since the wavelengths of
radiation corresponding to the frequencies of ion motion are much
larger than the trap structures, electric fields from blackbody
radiation (thermal electronic noise) are strongly altered from
their free-space values.  The fluctuating fields can be determined
by estimating the Johnson noise potential on each electrode from
electrode resistance and from resistors purposely attached to the
electrodes.  Here, we expect this resistance to be dominated by
that from the RC filters attached to each control electrode.  To
estimate the field at the position of the ion, we have numerically
solved for the potentials inside the trap structure.  For example,
for a 1 V potential applied to one of the $\#$1 electrodes we find
the axial field at the center of trap $\#$2 to be 2.42 V/cm. Using
this value and assuming the RC-filter resistors are at the ambient
temperature ($\sim$ 300 K), we estimate the axial heating to be
approximately 4 quanta per second for $\nu_z =$ 2.9 MHz.
Therefore the heating appears to be dominated by causes other than
thermal electronic noise \cite{turchette00}.

\section{Ion Transfer}
\noindent We transferred an ion from trap $\#$2 to trap $\#$4
(Fig. 1) and back by continuously changing the potentials on the
five pairs of control electrodes.  This translated the position
$z_0$ of the axial trap minimum between the two trap locations. We
initially prepared the ion in the $\ds | n_z = 0\rangle$ state,
where here we consider only the axial mode.  Using numerical
solutions for our trap geometry, trap potentials were designed so
that during the translation, $\nu_x$, $\nu_y$, and $\nu_z$ would
be held constant; for the experiments reported here, $\nu_z =$ 2.9
MHz. Starting at time $t=0$ in trap $\#$2 ($z_{0}=0$) the axial
trap position was smoothly translated according to
\begin{equation}
z_{0}(t)= sin^2\left(\frac{\pi t}{2 T} \right) \cdot 1.2 \,
\mathrm{mm}
\end{equation}
until time $t= T$, after which the axial trap remained at 1.2 mm,
the position of trap $\#$4. After a hold period approximately
equal to $T$ in trap $\#$4 the transfer process was reversed.
Following the transfer back to trap $\#$2 we measured the $\langle
n_z \rangle$ as described above.

To determine the motional heating due to the transfer we also
measured the motional state of the ion held in trap $\#$2 after
the same total time delay. Table 2 gives the results for the ion's
motional heating due to transfer back and forth (the heating in
the non-transfer case subtracted from the heating in the dynamic
case).
\begin{table}
\tcaption{The number of axial quanta gained, $\Delta n_z$, as a
result of transferring the ion from trap $\#$2 to trap $\#$4 and
back with a one-way transfer time interval $T$.  The axial trap
frequency, $\nu_z =$ 2.9 MHz.} \centerline{\footnotesize }
\centerline{\footnotesize\smalllineskip
\begin{tabular}{c c}\\
\hline
$T$ ($\mu$s)  & $\Delta n_z$  \\
\hline
16 & nonadiabatic  \\
28 & 0.6$\pm$0.2  \\
43 & 0.04$\pm$0.03  \\
54 & 0.01$\pm$0.03  \\
200 & 0.07$\pm$0.06  \\
300 & 0.00$\pm$0.07  \\
590 & 0.1$\pm$0.1  \\
\hline\\
\end{tabular}}
\end{table}
For $T=$ 16 $\mu$s, the measured motional sidebands were the same
size. This implies that very little population remained in the
motional ground state after transfer \cite{turchette00,monroe95}
and that the transfer was no longer adiabatic. For $T=$ 28 $\mu$s
about half a quantum on average was gained due to transfer. From a
numerical integration of the classical equations of motion we
expected that the ion should gain the amount of energy equal to
one motional quantum for a 30 $\mu$s transfer duration. This
estimate indicates approximately when the transfers are no longer
adiabatic and agrees reasonably well with our observations.
Overall, this transfer process is robust in that we have not
observed any ion loss due to transfer. As an example, in a series
of experiments for $T$ = 54 $\mu$s we transferred the same ion
over $10^{6}$ times. In addition, we measured the heating of the
two radial motional modes (frequencies between 4 and 5 MHz) due to
transfer. For $T$ = 54 $\mu$s the heating for both of these modes
was less than 1 quantum.

\section{Coherence Test}
\begin{figure} 
\vspace*{13pt}
\centerline{\psfig{file=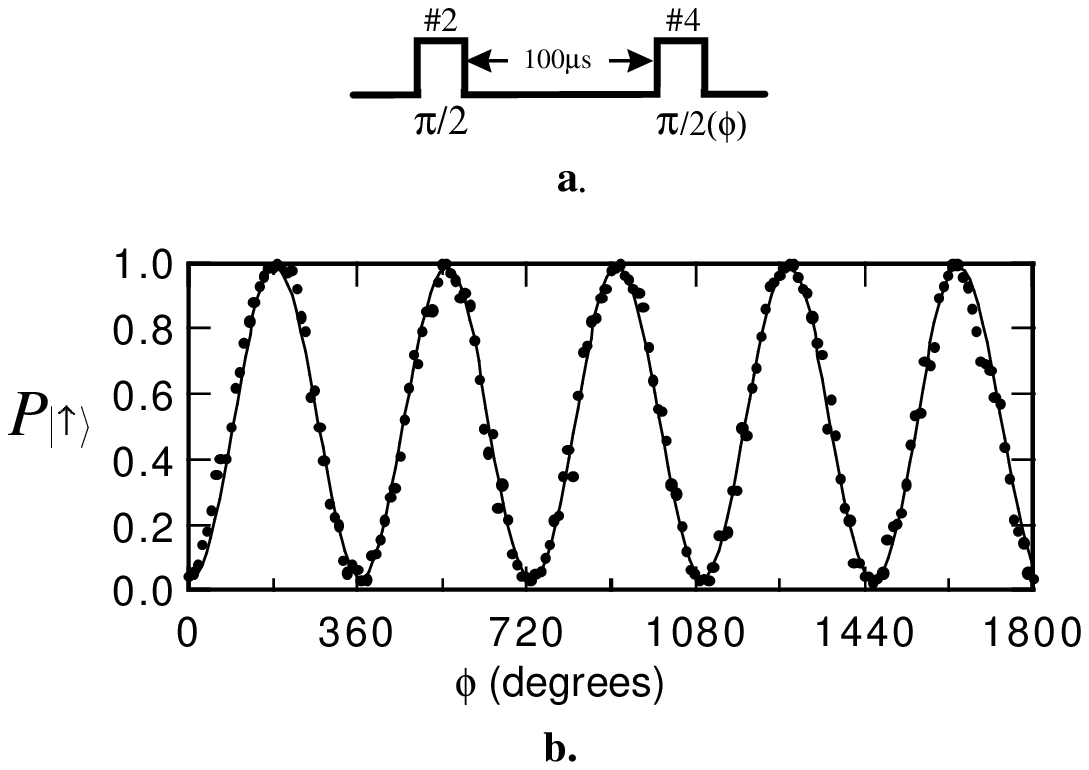, width=8.2cm}} 
\vspace*{13pt} \fcaption{\label{ramsey_fringes} Coherence transfer
experiment.  a. The experimental timing. b.  The probability for
the final state to be $\mid \uparrow \rangle$, $P_{\mid \uparrow
\rangle}$, as a function of the phase, $\phi$, of the final $\pi /
2$ pulse.  Each graphed point is the result of 200 individual
experiments.  The probability, $P_{\mid \uparrow \rangle}$, is the
fraction of these experiments with 3 or fewer photons detected
during the detection period.  (When the ion was in the
$|\downarrow \rangle$ state, the mean number of detected photons
was 12.)}
\end{figure}

\noindent A critical requirement for the viability of the
multiplexed system envisioned here is that the internal coherence
of the ions is maintained as they are moved.  We verified this
with a Ramsey-type interference experiment.  After laser cooling
to the $\ds |n_z = 0 \rangle$ state in trap $\#$2, we carried out
the transformation
\begin{equation}
\ds \rightarrow \frac{1}{\sqrt{2}} \left(\ds + \us \right)
\end{equation}
by applying a $\pi / 2$ pulse using the 0$^{\circ}$ geometry for
the Raman laser beams. We then transferred the ion from trap $\#2$
to trap $\#4$ in $T$ $\simeq$ 55 $\mu$s. The coherence of the
internal state remaining after transfer was measured with a final
Raman $\pi / 2$ pulse (0$^{\circ}$ geometry) in trap $\#4$ with a
controllable phase $\phi$ relative to the first pulse
\cite{wineland98}, which was varied. The ion was then moved back
to trap $\#$2 for state measurement. The probability $P_{\mid
\uparrow \rangle}$ of finding the final state of the ion to be
$\us$ is
\begin{equation}
P_{\mid \uparrow \rangle} = \frac{1}{2} \left( 1+C\cos (\phi +\phi ^{\prime}) \right)
\end{equation}
where $\phi ^{\prime}$ is a constant laser phase difference
between traps $\#$2 and $\#4$. The fringe contrast $C$ is a
measure of the coherence \cite{turchette00b}.
Figure~\ref{ramsey_fringes}a shows the timing of the overall
experiment.
The length of the $\pi / 2$ pulses was 1 to 2 $\mu$s.  The
interval between $\pi / 2$ pulses was 100 $\mu$s.
Figure~\ref{ramsey_fringes}b shows the oscillation in $P_{\mid
\uparrow \rangle}$ as $\phi$ was varied.  The fringe contrast was
$95.8\pm0.8\ \%$, indicating the preservation of coherence.  This
experiment was line-triggered (60 Hz) to minimize the loss of
contrast due to magnetic fields fluctuating at 60 Hz and harmonics
of 60 Hz. The data for Figure~\ref{ramsey_fringes}b were taken
with 37,000 consecutive round-trip transfers of the same ion.

In another experiment we used a spin-echo technique to minimize
the effect of fluctuating magnetic fields so we could trigger the
experiment without 60 Hz synchronization.  The qubit transition
frequency depends on magnetic field ($\simeq 2.1 \times 10^{10}$
Hz-T$^{-1}$), so the internal state of the ion accumulates an
uncontrolled phase between $\us$ and $\ds$ during the
free-evolution period between $\pi / 2$ pulses due to a
time-varying ambient magnetic field. This uncontrolled phase
causes the Ramsey fringes to partially wash out after averaging
over many experiments, but since the magnetic field changes
negligibly during the time of a single experiment we are able to
correct for it. By inserting a $\pi$ pulse between the two Ramsey
pulses the phase accumulated during the first half of the Ramsey
period is cancelled by that accumulated during the second half.

\begin{figure} 
\vspace*{13pt}
\centerline{\psfig{file=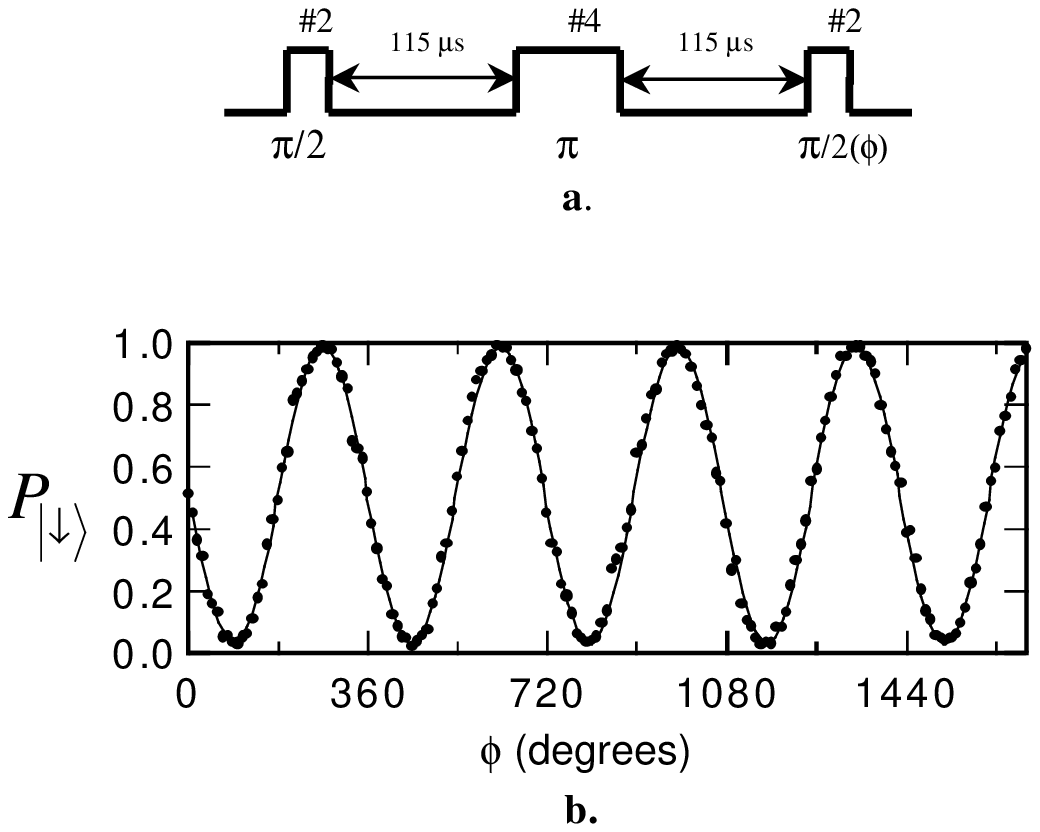, width=8.2cm}} 
\vspace*{13pt} \fcaption{\label{bang_bang_expt} Spin-echo
coherence transfer experiment.  a. The experimental timing. b.
The probability for the final state to be $\mid \uparrow \rangle$,
$P_{\mid \uparrow \rangle}$, as a function of the phase, $\phi$,
of the final $\pi / 2$ pulse.  Each graphed point is the result of
1000 individual experiments.  The probability, $P_{\mid \uparrow
\rangle}$, is the fraction of these experiments with 3 or fewer
photons detected during the detection period. (When the ion was in
the $|\downarrow \rangle$ state, the mean number of detected
photons was 12.)  The two $\pi/2$ pulses were generated from
separate sources which had a constant phase offset; therefore, a
minimum does not occur for $\phi =$ 0 as would otherwise be
expected.}
\end{figure}

Figure~\ref{bang_bang_expt}a shows the timing for this experiment.
Here, we applied both $\pi / 2$ pulses in trap $\#$2.  Between
these $\pi/2$ pulses we transferred the ion to trap $\#$4  where
we applied the $\pi$ pulse. Figure~\ref{bang_bang_expt}b shows the
oscillation in $P_{\mid \uparrow \rangle}$ as the phase of the
final $\pi / 2$ pulse, $\phi$, was scanned.  The fringe contrast
is $96.6\pm 0.5\ \%$. These data were taken with 170,000
consecutive round-trip transfers of the same ion.  In a control
experiment we applied the same pulses and timing but did not
transfer the ion (the $\pi$ pulse was applied in trap $\#$2).  For
these data we measured a fringe contrast equal to $96.8\pm 0.3\ \%
$. The reduction in fringe contrast from 100$\ \% $ in the control
experiment is due to factors other than transfer decoherence, such
as imperfect Raman pulses and imperfect state preparation and
detection.  The decrease in fringe contrast due to transferring
the ion both ways was $0.2 \pm 0.6\ \% $. Therefore, within
experimental error, we see no reduction in the internal state
coherence of the ion due to transfer.

\section{Separating Ions}
\noindent As a test of separating ions, we confined two ions in
trap $\#$2, separated them into traps $\#$2 and $\#$4 (1.2 mm
apart), and then brought them back together in trap $\#$2. The
separation sequence began with laser cooling in trap $\#$2 after
which the two ions were transferred from trap $\#$2 to trap $\#$3.
Prior to separation, the axial center-of-mass frequency in trap
$\#$3 was adjusted to 700 kHz (8V, 0V, 0V, 0V, 8V on electrodes 1
through 5 respectively). Following this, the potential on
electrodes 3 was raised. Initially this process weakened the axial
confinement common to both ions. The trap potentials were smoothly
varied so that after approximately 2 ms, the lowest center-of-mass
axial frequency (90 kHz) was reached, at which point two separate
potential wells developed with one ion in each well.\footnote{\ \
For a single ion subjected to the same process, the axial
frequency goes to zero at the center of the trap before the double
well forms. However for two ions held apart by the Coulomb
interaction, neither the center-of-mass nor symmetric-stretch mode
frequencies go through zero.} \ \  From this point, the individual
ion axial frequencies were smoothly increased to about 2.9 MHz.
During the separation, the $x$ and $y$ mode frequencies smoothly
varied and were bounded between 3 and 7 MHz. We found that by
raising the potentials on electrodes 1 and 5 in addition to
electrodes 3 during separation, we had finer control over the ion
motion. The potential barriers from electrodes 1 and 5 prevented
the ions from ``rolling down" the potential hill created by
electrodes 3, making the process less sensitive to voltage
fluctuations on electrodes 3. Even with this strategy we had to
raise the potential on electrodes 3 in a precisely controlled
manner.  For example, the finite voltage steps ($\approx$ 10 mV)
in the output of the digital waveform generators applied to the
electrodes heated the ions.  We filtered these potential steps
with an additional cascade of two $RC$ low-pass filters ($R$ = 1
k$\Omega$, $C$ = 22 nF) inserted before the control electrode
filters shown in Fig. 1d. When the ions were finally separated
into traps $\#$2 and $\#$4, the voltages on the electrodes were
respectively 8V, 0V, 8V, 0V, 8V on electrodes 1 through 5. The
axial frequency $\nu_z$ was equal to 2.9 MHz in each of these
traps. In the separation process, the trajectories of the ions
were designed to start and end with zero velocity and
acceleration; the acceleration between these end points was a
(smooth) polynomial function. In the experiments reported here, we
used a total period of $\approx$ 10 ms to separate the ions from
trap $\#$3 to traps $\#$ 2 and 4. This process was reversed to
bring the ions back together in trap $\#$2.

\begin{figure} 
\vspace*{13pt}
\centerline{\psfig{file=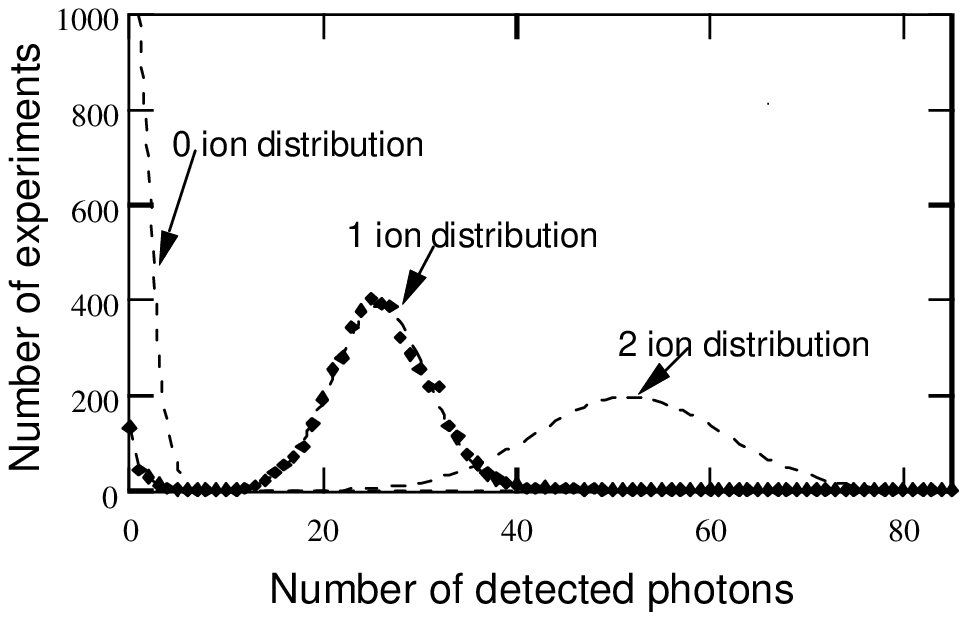, width=8.2cm}} 
\vspace*{13pt} \fcaption{\label{separating} Histogram of the
number of photons detected in a $200\ \mu$s time interval for 5050
separation experiments. The distributions of the photon counts for
0, 1, and 2 ions detected are shown as well.  The 0-ion
distribution extends to 3030 for the 0-photon bin.}
\end{figure}

In order to demonstrate separation of the ions, we measured the
number of ions in trap $\#$2 during the period when the ions were
supposed to be separated into trap $\#$2 and trap $\#$4.  Our
detection optics collected fluorescence photons only from trap
$\#$2.  After applying Doppler cooling to any ions present in trap
$\#$2 for 2 ms we probed them with the detection and repump
lasers. This determined the number of ions present in trap $\#$2
independent of their internal state.  Figure~\ref{separating}
shows a histogram resulting from a run consisting of 5050
consecutive separation experiments.  The horizontal axis is the
number of detected photons recorded in an experiment; the vertical
axis displays the number of experiments in which that number was
detected.
The distributions (0, 1, and 2 ion) for different ion numbers are
shown for reference.  By fitting the data to these distributions
we determined that in $95\pm1\ \%$ of the experiments one ion was
in trap $\#$2.  In $4\pm1\%$ of the experiments no ions were in
trap $\#$2 and in less than $1\%$ of the experiments two ions were
in trap $\#$2.  We found that we could pick off any number of ions
(0, 1, or 2) and move them to trap $\# 2$ by changing the bias
potential between electrodes 1 and 5. A change of bias on the
order of 25 mV adjusted the ion split. The asymmetry observed for
the data of Fig. 4 resulted from our inability to precisely
control this offset.  Finally, with three ions we saw the photon
levels corresponding to 0, 1, 2, or 3 ions in trap $\#2$ as we
changed the bias between electrodes 1 and 5.

\begin{figure} 
\vspace*{13pt}
\centerline{\psfig{file=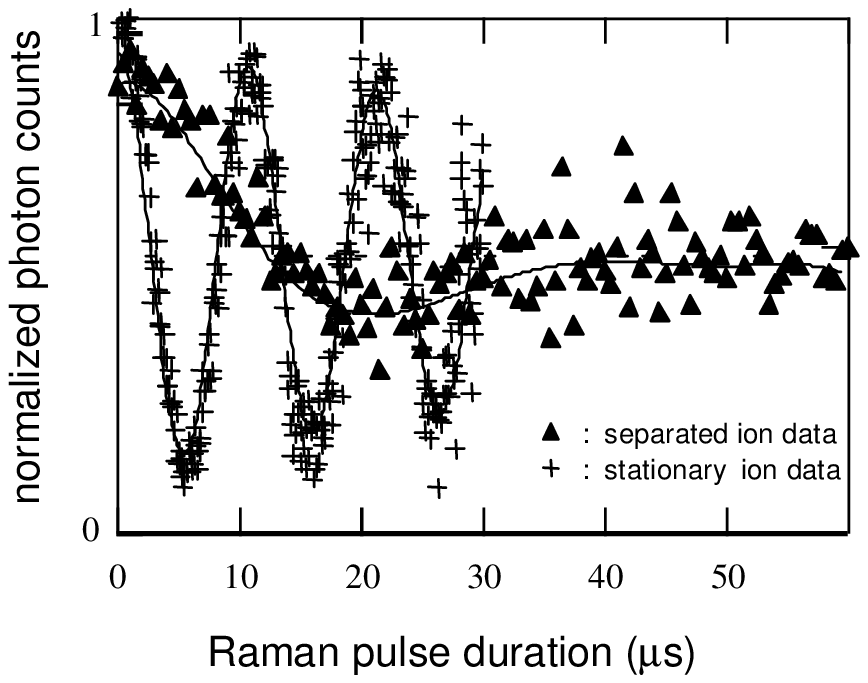, width=8.2cm}} 
\vspace*{13pt} \fcaption{\label{sep_heat} Photon counts detected
as a function of the Raman (carrier) pulse duration.  Data for a
``hot" separated ion and a single cold ion that has not been moved
are shown. The separated-ion data are connected with a smooth
curve to guide the eye.  The cold ion data are fit with the
expected curve to give $\langle n_z \rangle =0.16$.}
\end{figure}

We also measured the energy gained by an ion due to the separating
process.  Because the ions gained many quanta, the
sideband-comparison measurement technique was no longer sensitive.
Instead we used stimulated-Raman carrier ($\ds |n_z \rangle
\leftrightarrow \us |n_z \rangle $) transitions in the $|n_z
\rangle$-state sensitive $90^{\circ}$ geometry with $\Delta
\vec{k}$ aligned along the trap axis \cite{wineland98,nagerl00}.
After separating the two ions into trap $\#$2 and trap $\#$4 we
drove Raman carrier transitions on the ion in trap $\#$2.  The
final state population of the ion in trap $\#$2 was measured with
a standard detection pulse. Figure~\ref{sep_heat} shows the
oscillation in detected photon counts (proportional to the final
state population) as a function of the Raman pulse duration.
The populations of different $|n_z \rangle$-states oscillate at
different rates \cite{wineland98,nagerl00} so the state population
for the ``hot" separated ion (a state with a distribution of many
$|n_z \rangle$-states for an ensemble of experiments) made about
half an oscillation before the different $|n_z \rangle$ state
oscillation periods averaged to the midway level (half $\us$ and
half $\ds$). Also shown in Figure~\ref{sep_heat} are data from a
single ion prepared in $\ds $ and cooled to near the ground state
in trap $\#$2. By comparing the time of the first midway level
crossing in the figure for the separated-ion data to that for the
``cold" ion data we estimated the average value of $n_z$ for the
separated ion. Assuming a thermal distribution of $n_z$ levels we
found the average $\langle n_z \rangle$ for the separated ion to
be 140 $\pm$ 70 quanta.

The 10 ms separation time minimized the observed heating during
separation.  At the present time, we do not understand why this
time could not be shorter. However, we did observe that most of
the energy increase occurred when the center-of-mass frequency was
smallest.  This was verified by allowing the separation process to
evolve only to a certain point, reversing it, and then measuring
the ions' temperature. In this way we have an approximate measure
of the heating integrated up to the time of reversal. This
measurement revealed that the energy increased sharply near the
point where the center-of-mass frequency was smallest. This
increase in heating at lower frequencies can be expected for two
reasons.  First, the heating (in terms of quanta per unit of time)
increases as $\nu_z^{-1}$ even for a uniform noise spectrum.
Moreover, Johnson noise from the $RC$ filters and any noise
injected externally on the control electrodes will increase
substantially at lower frequencies due to the reduced efficiency
of the filters. This was supported by experiments where we
measured the single-ion heating at axial frequencies below 1 MHz,
observing that the heating increased more strongly than
$\nu_z^{-3}$.  Future studies will have to address the causes of
heating at lower frequencies and its dependence on $\nu_z$.

\section{Summary and Conclusions}
\noindent In these experiments, we have demonstrated some of the
key elements for scaling up an ion-trap quantum processor based on
the idea of moving ion qubits between nodes of a multiplexed trap
array \cite{wineland98,kielpinski02}. We have adiabatically
transferred (no motional heating) an ion between two distinct
traps, 1.2 mm apart, in approximately 50 $\mu$s.  We have
transferred a single ion between these traps more than $10^{6}$
consecutive times and have observed no loss due to transfer.  We
measured the fringe contrast in a Ramsey-type experiment where an
ion was transferred back and forth between two traps; the measured
loss of contrast from transfer was $0.2\pm0.6\ \%$. Therefore,
within experimental error, the internal state coherence of the ion
was preserved during transfer.  In addition, we demonstrated the
separation of two ions, initially held in one trap, into two
distinct traps 1.2 mm apart.  The separation process took 10 ms
and was successful $95\%$ of the time.  We measured the energy
gained by the ions due to separation to be approximately 150
quanta assuming a thermal distribution of the final motional
levels. Finally, we have seen a reduction in the motional heating
compared to previous traps \cite{turchette00} by a factor of
approximately 100.   This improvement appears to be related to the
integrity of the trap electrode surfaces and may have been due to
preventing deposition of beryllium on the electrodes and/or
insuring more complete coating of the electrodes substrates with
gold.

The heating that was observed upon separating ions was not
unanticipated; to combat it, we expect that some sort of
sympathetic cooling must be employed \cite{wineland98}.  On the
other hand, we believe that it can be reduced and the separation
time decreased substantially in future experiments.  For example,
the geometry of this trap is not well-suited for separating ions.
At the minimum center-of-mass axial trap frequency during
separating ($\sim$90 kHz) the ions are separated by about 50
$\mu$m.  Yet, in the experiments reported here, we effectively
insert an approximately 800 $\mu$m wide potential wedge between
them at this point, making separation very sensitive to small
field offsets. If we employ electrode dimensions where the
distance from the ion to the nearest electrode is about 50 $\mu$m,
this should make the width of wedge approximately equal to this
distance and make electrode voltage control much less stringent.
Of course, in the past \cite{turchette00}, motional heating has
been observed to increase significantly as the trap dimensions are
reduced. However, by taking further steps to insure the integrity
of the electrode surfaces, we believe that making smaller traps
can lead to efficient separation without significant heating.  To
minimize heating that might result from smaller dimensions, we
note that the separation could be performed in specific locations
of the trap array where the dimensions are much smaller than in
the accumulator and storage regions.

Optical forces might also be used to facilitate separation (and
transfer) as has been demonstrated for neutral atoms
\cite{schrader01,gustavson02}.  A high-intensity, far-detuned
focused laser beam could be applied to the logical qubit ions to
exert an optical-dipole force without causing significant
dephasing or decoherence due to spontaneous emission.  An
alternative approach would be to apply optical forces to the
cooling ions used for sympathetic cooling.  One version of this
idea would be to make a logical qubit ``packet" where two cooling
ions would surround each physical qubit or each logical qubit
encoded into a set of physical qubits.  Each packet would be kept
intact as it is moved through the device by applying optical
forces to the cooling ions.  This would have the advantage that
near-resonant optical-dipole or spontaneous scattering forces
could be applied to the cooling ions thereby reducing the laser
intensity required for a given force.  Although spontaneous
emission of the cooling ions would be relatively high, it would
not significantly affect the qubit ions, particularly if the qubit
and cooling ion transition frequencies are substantially
different. When separation with minimal heating is achieved,
exciting possibilities such as complex entanglement studies,
repetitive error correction, and quantum computing may become
practical.


\nonumsection{Acknowledgements} \noindent  This work was supported
by the U. S. National Security Agency (NSA) and Advanced Research
and Development Activity (ARDA) under Contract No. MOD-7171.00,
the U. S. Office of Naval Research (ONR), and the U. S. National
Reconnaissance Office (NRO).  We thank M. Barrett, D. Lucas, D.
Smith, and D. Sullivan for helpful comments on the manuscript.

\nonumsection{References}


\noindent

\begin{thebibliography}{000}
\bibitem{lloyd93}
S. Lloyd (1993), {\it A potentially realizable quantum computer},
Science {\bf 261}, pp. 1569-1571.

\bibitem{cirac97}
J. I. Cirac, P. Zoller, H. J. Kimble, and H. Mabuchi (1997), {\it
Quantum State Transfer and Entanglement Distribution among Distant
Nodes in a Quantum Network} Phys. Rev. Lett. {\bf 78}, pp.
3221-3224.

\bibitem{pellizzari97}
T. Pellizzari (1997), {\it Quantum networking with optical
fibers}, Phys. Rev. Lett. {\bf 79}, pp. 5242-5245.

\bibitem{deVoe98}
R. G. DeVoe (1998), {\it Elliptical ion traps and trap arrays for
quantum computation}, Phys. Rev. A {\bf 58}, pp. 910-914.

\bibitem{kikkawa99}
J. M. Kikkawa and D. D. Awschalom (1999), {\it Lateral drag of
spin coherence in gallium arsenide}, Nature {\bf 397}, pp.
139-141.

\bibitem{barnes00}
C. Barnes, J. M. Shilton, and A. M. Robinson (2000), {\it Quantum
computation using electrons trapped by surface acoustic waves},
Phys. Rev. B {\bf 62}, pp. 8410-8419.

\bibitem{recher01}
P. Recher, E. V. Sukhhorukov, and D. Loss (2001), {\it Andreev
tunneling, Coulomb blockade, and resonant transport of nonlocal
spin-entangled electrons}, Phys. Rev. B {\bf 63}, pp. 165314-1-11.

\bibitem{wineland98}
D. J. Wineland, C. Monroe, W. M. Itano, D. Leibfried, B. E. King,
and D. M. Meekhof (1998), {\it Experimental issues in coherent
quantum-state manipulation of trapped atomic ions}, J. Res. Natl.
Inst. Stand. Technol. {\bf 103}, pp. 259-328.

\bibitem{cirac00}
J. I. Cirac and P. Zoller (2000), {\it A scalable quantum computer
with ions in an array of microtraps}, Nature {\bf 404}, pp.
579-581.

\bibitem{renn95}
M. J. Renn, D. Montgomery, O. Vdovin, D. Z. Anderson, C. E.
Wieman, and E. A. Cornell (1995), {\it Laser-guided atoms in
hollow-core optical fibers}, Phys. Rev. Lett. {\bf 75}, pp.
3253-3256.

\bibitem{denschlag99}
J. Denschlag, D. Cassettari, and J. Schmiedmayer (1999), {\it
Guiding neutral atoms with a wire}, Phys. Rev. Lett. {\bf 82}, pp.
2014-2017.

\bibitem{muller99}
D. M{\"u}ller, D. Z. Anderson, R. J. Grow, P. D. D. Schwindt, and
E. A. Cornell (1999), {\it Guiding neutral atoms around curves
with lithographically patterned current-carrying wires}, Phys.
Rev. Lett. {\bf 83}, pp. 5194-5197.

\bibitem{dekker00}
N. H. Dekker, C. S. Lee, V. Lorent, J. H. Thywissen, S. P. Smith,
M. Drndi, R. M. Westervelt, and M. Prentiss (2000), {\it Guiding
neutral atoms on a chip}, Phys. Rev. Lett. {\bf 84}, pp.
1124-1127.

\bibitem{fortagh00}
J. Fortagh, H. Ott, A. Grossmann, C. Zimmermann (2000), {\it
Miniaturized magnetic guide for neutral atoms}, Appl. Phys. B {\bf
70}, pp. 701-708.

\bibitem{key00}
M. Key, I. G. Hughes, W. Rooijakkers, B. E. Sauer, and E. A. Hinds
(2000), {\it Propagation of cold atoms along a miniature magnetic
guide}, Phys. Rev. Lett. {\bf 84}, pp. 1371-1373.

\bibitem{cassettari00}
D. Cassettari, A. Chenet, R. Folman, A. Haase, B. Hessmo, P.
Kr{\"u}ger, T. Maier, S. Schneider, T. Calarco, J. Schmiedmayer
(2000), {\it Micromanipulation of neutral atoms with
nanofabricated structures}, Appl. Phys. B {\bf 70}, pp. 721-730.

\bibitem{muller01}
Dirk M\"{u}ller, Eric A. Cornell, Marco Prevedelli, Peter D. D.
Schwindt, Ying-Ju Wang, and Dana Z. Anderson (2001), {\it Magnetic
switch for integrated atom optics}, Phys. Rev. A. {\bf 63},
041602(R).

\bibitem{reichel99}
J. Reichel, W. H\"{a}nsel, and T. W. H\"{a}nsch  (1999), {\it
Atomic micromanipulation with magnetic surface traps}, Phys. Rev.
Lett. {\bf 83}, pp. 3398-3401.

\bibitem{schrader01}
D. Schrader, S. Kuhr, W. Alt, M. M{\"u}ller, V. Gomer, D. Meschede
(2001), {\it An optical conveyor belt for single neutral atoms},
Appl. Phys. B {\bf 73}, pp. 819 -824.

\bibitem{gustavson02}
T. L. Gustavson, A. P. Chikkatur, A. E. Leanhardt, A. G\"{o}rlitz,
S. Gupta, D. E. Pritchard, and W. Ketterle  (2002), {\it Transport
of Bose-Einstein condensates with optical tweezers}, Phys. Rev.
Lett. {\bf 88}, 020401.

\bibitem{prestage94}
J. D. Prestage, R. L. Tjoelker, G. J. Dick, and L. Maleki (1994),
{\it Improved linear ion trap physics package}, Proc. 1993 IEEE
Frequency Control Symposium, pp. 144-147. and J. D. Prestage, R.
L. Tjoelker, and L. Maleki (2001), {\it Recent developments in
microwave ion clocks}, Topics Appl. Phys. {\bf 79}, pp. 195-211.

\bibitem{guthohrlein01}
G. R. Guth\"{o}hrlein, M. Keller, K. Hayasaka, W. Lange, and H.
Walther (2001), {\it A single ion as a nanoscopic probe of an
optical field}, Nature {\bf 414}, pp. 49-51.

\bibitem{cirac95}
J. I. Cirac and P. Zoller (1995), {\it Quantum computations with
cold trapped ions}, Phys. Rev. Lett. {\bf 74}, pp. 4091-4094.

\bibitem{steane00}
A. Steane, C. F. Roos, D. Stevens, A. Mundt, D. Leibfried, F.
Schmidt-Kaler, and R. Blatt (2000), {\it Speed of ion-trap
quantum-information processors} Phys. Rev. A {\bf 62}, pp.
042305-1-9.

\bibitem{enzer00}
D. G. Enzer, M. M. Schauer, J. J. Gomez, M. S. Gulley, M. H.
Holzscheiter, P. G. Kwiat, S. K. Lamoreaux, C. G. Peterson, V. D.
Sandberg, D. Tupa, A. G. White, and R. J. Hughes (2000), {\it
Observation of power-law scaling for phase transitions in linear
ion crystals}, Phys. Rev. Lett. {\bf 85}, pp. 2466-2469.

\bibitem{kielpinski02}
D. Kielpinski, C. Monroe, and D. J. Wineland (2002), {\it
Architecture for a large-scale ion-trap quantum computer},
submitted to {\it Nature}.

\bibitem{turchette00}
Q. A. Turchette, D. Kielpinski, B. E. King, D. Leibfried, D. M.
Meekhof, C. J. Myatt, M. A. Rowe, C. A. Sackett, C. S. Wood, W. M.
Itano, C. Monroe, and D. J. Wineland (2000), {\it Heating of
trapped ions from the quantum ground state}, Phys. Rev. A {\bf
61}, pp. 063418-1-8.

\bibitem{berkeland98}
D. J. Berkeland, J. D. Miller, J. C. Bergquist, W. M. Itano, and
D. J. Wineland (1998), {\it Minimization of ion micromotion in a
Paul trap}, J. Appl. Phys. {\bf 83}, pp. 5025-5033.

\bibitem{jefferts95}
S. R. Jefferts, C. Monroe, E. W. Bell, and D. J. Wineland (1995),
{\it Coaxial-resonator-driven RF (Paul) trap for strong
confinement}, Phys. Rev. A {\bf 51}, pp. 3112-3116.

\bibitem{monroe95}
C. Monroe, D. M. Meekhof, B. E. King, W. M. Itano, and D. J.
Wineland (1995), {\it Demonstration of a fundamental quantum logic
gate}, Phys. Rev. Lett. {\bf 75}, pp. 4714-4717.

\bibitem{king98}
B. E. King, C. S. Wood, C. J. Myatt, Q. A. Turchette, D.
Leibfried, W. M. Itano, C. Monroe, and D. J. Wineland (1998), {\it
Cooling the collective motion of trapped ions to initialize a
quantum register}, Phys. Rev. Lett. {\bf 81}, pp. 1525-1528.

\bibitem{meekhof96}
D. M. Meekhof, C. Monroe, B. E. King, W. M. Itano, and D. J.
Wineland (1996), {\it Generation of nonclassical motional states
of a trapped atom}, Phys. Rev. Lett. {\bf 76}, pp. 1796-1799.

\bibitem{sackett01}
C. A. Sackett (2001), {\it Quantum information experiments with
trapped ions: status and prospects}, Quant. Inf. Comp. {\bf 1},
pp. 57-80.

\bibitem{rowe01}
M. A. Rowe, D. Kielpinski, V. Meyer, C. A. Sackett, W. M. Itano,
C. Monroe, and D. J. Wineland (2001), {\it Experimental violation
of a Bell's inequality with efficient detection}, Nature {\bf
409}, pp. 791-794.

\bibitem{Rohde01}
H. Rohde, S. T. Gulde, C. F. Roos, P. A. Barton, D. Leibfried, J.
Eschner, F. Schmidt-Kaler, and R. Blatt (2001), {Sympathetic
ground-state cooling and coherent manipulation with two-ion
crystals}, J. Opt. B: Quantum Semiclass. Opt. {\bf 3} pp. S34 -
S41.

\bibitem{turchette00b}
Q. A. Turchette, C. J. Myatt, B. E. King, C. A. Sackett, D.
Kielpinski, W. M. Itano, C. Monroe, and D. J. Wineland (2000),
{\it Decoherence and decay of motional quantum states of a trapped
atom coupled to engineered reservoirs}, Phys. Rev. A {\bf 62}, pp.
053807-1-22.


\bibitem{nagerl00}
H. C. N\"{a}gerl, Ch. Roos, D. Leibfried, H. Rohde, G. Thalhammer,
J. Eschner, F. Schmidt-Kaler, and R. Blatt (2000), {\it
Investigating a qubit candidate: Spectroscopy on the $S_{1/2}$ to
$D_{5/2}$ transition of a trapped calcium ion in a linear Paul
trap}, Phys. Rev. A {\bf 61}, 023405-1-9.

\end{thebibliography}
\end{document}